\documentclass[aps,prb,twocolumn,superscriptaddress,showpacs,showkeys]{revtex4-2}

\usepackage{color}
\usepackage{graphicx}
\usepackage{dcolumn}
\usepackage{bm}
\usepackage{float}
\usepackage[mathlines]{lineno}
\usepackage{tabularx}
\usepackage[colorlinks,linkcolor=blue,anchorcolor=blue,citecolor=blue,urlcolor=blue,hyperindex,CJKbookmarks]{hyperref}
\usepackage{footnote}
\usepackage{amsmath}
\usepackage{txfonts}
\usepackage{mathptmx}
\usepackage{ragged2e}
\usepackage{booktabs,makecell, multirow, tabularx}
\usepackage{multirow}
\usepackage{braket}
\usepackage{gensymb}
\usepackage{array}

\newcolumntype{L}[1]{>{\raggedright\let\newline\\\arraybackslash\hspace{0pt}}m{#1}}
\newcolumntype{C}[1]{>{\centering\let\newline\\\arraybackslash\hspace{0pt}}m{#1}}
\newcolumntype{R}[1]{>{\raggedleft\let\newline\\\arraybackslash\hspace{0pt}}m{#1}}

\begin{document}

\title{DFT+DMFT study of correlated electronic structure in the monolayer-trilayer phase of La$_3$Ni$_2$O$_7$}

\author{Zhenfeng Ouyang}\affiliation{Department of Physics and Beijing Key Laboratory of Opto-electronic Functional Materials $\&$ Micro-nano Devices, Renmin University of China, Beijing 100872, China}\affiliation{Key Laboratory of Quantum State Construction and Manipulation (Ministry of Education), Renmin University of China, Beijing 100872, China}
\author{Jia-Ming Wang}\affiliation{Department of Physics and Beijing Key Laboratory of Opto-electronic Functional Materials $\&$ Micro-nano Devices, Renmin University of China, Beijing 100872, China}\affiliation{Key Laboratory of Quantum State Construction and Manipulation (Ministry of Education), Renmin University of China, Beijing 100872, China}
\author{Rong-Qiang He}\email{rqhe@ruc.edu.cn}\affiliation{Department of Physics and Beijing Key Laboratory of Opto-electronic Functional Materials $\&$ Micro-nano Devices, Renmin University of China, Beijing 100872, China}\affiliation{Key Laboratory of Quantum State Construction and Manipulation (Ministry of Education), Renmin University of China, Beijing 100872, China}
\author{Zhong-Yi Lu}\email{zlu@ruc.edu.cn}\affiliation{Department of Physics and Beijing Key Laboratory of Opto-electronic Functional Materials $\&$ Micro-nano Devices, Renmin University of China, Beijing 100872, China}\affiliation{Key Laboratory of Quantum State Construction and Manipulation (Ministry of Education), Renmin University of China, Beijing 100872, China}\affiliation{Hefei National Laboratory, Hefei 230088, China}

\date{\today}

\begin{abstract}
By preforming DFT+DMFT calculations, we systematically investigate the correlated electronic structure in the newly discovered monolayer-trilayer (ML-TL) phase of La$_3$Ni$_2$O$_7$ (1313-La327). Our calculated Fermi surfaces are in good agreement with the result of angle-resolved photoemission spectroscopy. We find that 1313-La327 is a multiorbital correlated metal. An orbital-selective Mott behavior is found in ML in our zero- and finite-temperature calculations. The ML Ni-3$d_{z^2}$ orbital exhibits a Mott behavior, while the ML Ni-3$d_{x^2-y^2}$ orbital is metallic due to self-doping. We also find a large static local spin susceptibility of ML Ni, suggesting that there is large spin fluctuation in 1313-La327. The TL Ni-$e_g$ orbitals possess similar electronic correlation to those in La$_4$Ni$_3$O$_{10}$. The $e_g$ orbitals of the outer-layer Ni in TL show non-Fermi liquid behaviors. Besides, large weight of high-spin states are found in TL-outer Ni and ML Ni, implying Hundness. Under 16 GPa,  a Lifshitz transition is revealed by our calculations and a La-related band crosses the Fermi level. Our work provides a theoretical reference for studying other potential mixed-stacked nickelate superconductors.  
\end{abstract}


\maketitle

\section{Introduction}
The discovery of superconductivity with transition temperature ($T_c$) about 80 K in Ruddlesden-Ropper (RP) bilayer perovskite La$_3$Ni$_2$O$_7$ (2222-La327) under pressure~\cite{Sun-nature} has attracted extensive attention. Before that, superconductivity in nickelates was found in doped infinite-layer R$_{0.8}$Sr$_{0.2}$NiO$_2$ thin films (R = La, Nd, and Pr) and quintuple-layer square-planar Nd$_6$Ni$_5$O$_{12}$ ~\cite{Li-nature572,doi:10.1021/acs.nanolett.0c01392,WOS:000721415700001}. Differed from the nominal 3$d^9$ electronic configuration in infinite-layer RNiO$_2$, RP series of nickelates La$_{n+1}$Ni$_n$O$_{3n+1}$ have higher chemical valence. With a chemical valence of Ni$^{2+}$, La$_2$NiO$_4$ ($n = 1$)~\cite{Rao-JSSC1984} is isostructural to superconducting La$_{2-x}$Sr$_x$CuO$_4$ and was once regarded as a platform to reproduce high-$T_c$ superconductivity like cuprates. The undoped La$_2$NiO$_4$ is an antiferromagnetic (AFM) insulator. With hole doping, the system shows a transition from AFM insulator to metal~\cite{Cava-prb1991,Chen-prl1993,Cheong-prb1994,Wu-prb2003}. However, no signatures of superconductivity were observed. 

For $n = 2$, with a higher average valence of Ni$^{2.5+}$, superconductivity transition was observed in 2222-La327 above 15 GPa~\cite{Sun-nature}. The x-ray-diffraction patterns show that the space group of the high-pressurized superconducting 2222-La327 is $I4/mmm$ with $180^{\degree}$ Ni-O-Ni bonding angle along $c$ axis~\cite{Li-arXiv2024,Wang-JACS2024}. Although a magnetic long-range order has not been experimentally observed, some theoretical research suggested that there are interlayer AFM superexchange interactions between the half-filled 3$d_{z^2}$ orbitals~\cite{Chen-prl2024,Tian-prb2024}. Our previous DFT+DMFT calculation~\cite{Ouyang-prb2024} has confirmed that the normal state of 2222-La327 is a Hund metal. The interlayer AFM superexchange interactions can be transmitted to the 3$d_{x^2-y^2}$ orbitals by Hund's coupling and then provide glue for superconducting pairing~\cite{Chen-prl2024,Tian-prb2024,Oh-prb2023}. Both model studies of functional renormalization group method~\cite{Yang-PRB2023} and cluster dynamical mean-field theory~\cite{Tian-prb2024} predicted an $s_{\pm}$-wave pairing.

Soon after, superconductivity transition in trilayer ($n = 3$) La$_4$Ni$_3$O$_{10}$ (La4310) was experimentally reported~\cite{Zhu-arXiv,Li-SCPMA67,Li-cpl2024,zhang2024superconductivity}. The maximal superconducting $T_c$ of La4310 is $\sim$ 20 K under high pressure. A structural transition from monolinic $P21/a$ to tetragonal $I4/mmm$ space group happens above 12 GPa~\cite{Li-SCPMA67}. Similar to 2222-La327, superconducting phase only emerges in high pressure phase without Ni-O octahedra distortions. Some theoretical studies also predicted a possible $s_{\pm}$-wave pairing~\cite{yang2-arXiv,zhang2-arXiv,zhang2024spmwave}. Our previous DFT+DMFT study~\cite{Wang-prb2024} suggests that inner-layer Ni in La4310 has a higher chemical valence than that of outer-layer Ni, which leads to weak electronic correlation in the inner layer. This may explain the lower superconducting $T_c$ of La4310. Another theoretical work~\cite{qin2024frustratedsuperconductivitytrilayernickelate} reveals intrinsic frustration in the spin-singlet pairing that the inner layer tends to form with both of the two outer layers respectively and thus may explain the reduction of $T_c$. According to Ref.~\cite{lu2024superconductivity}, the occupations of the Ni-3$d_{z^2}$ orbitals in La4310 are not half-filled, which weakens the effective interlayer couplings and causes the lower superconducting $T_c$ than that of 2222-La327. It seems that RP bilayer structure becomes the most promising candidate for realizing high-$T_c$ superconductivity in the family of nickelates.

\begin{figure*}[thb]
\centering
\includegraphics[width=17.2cm]{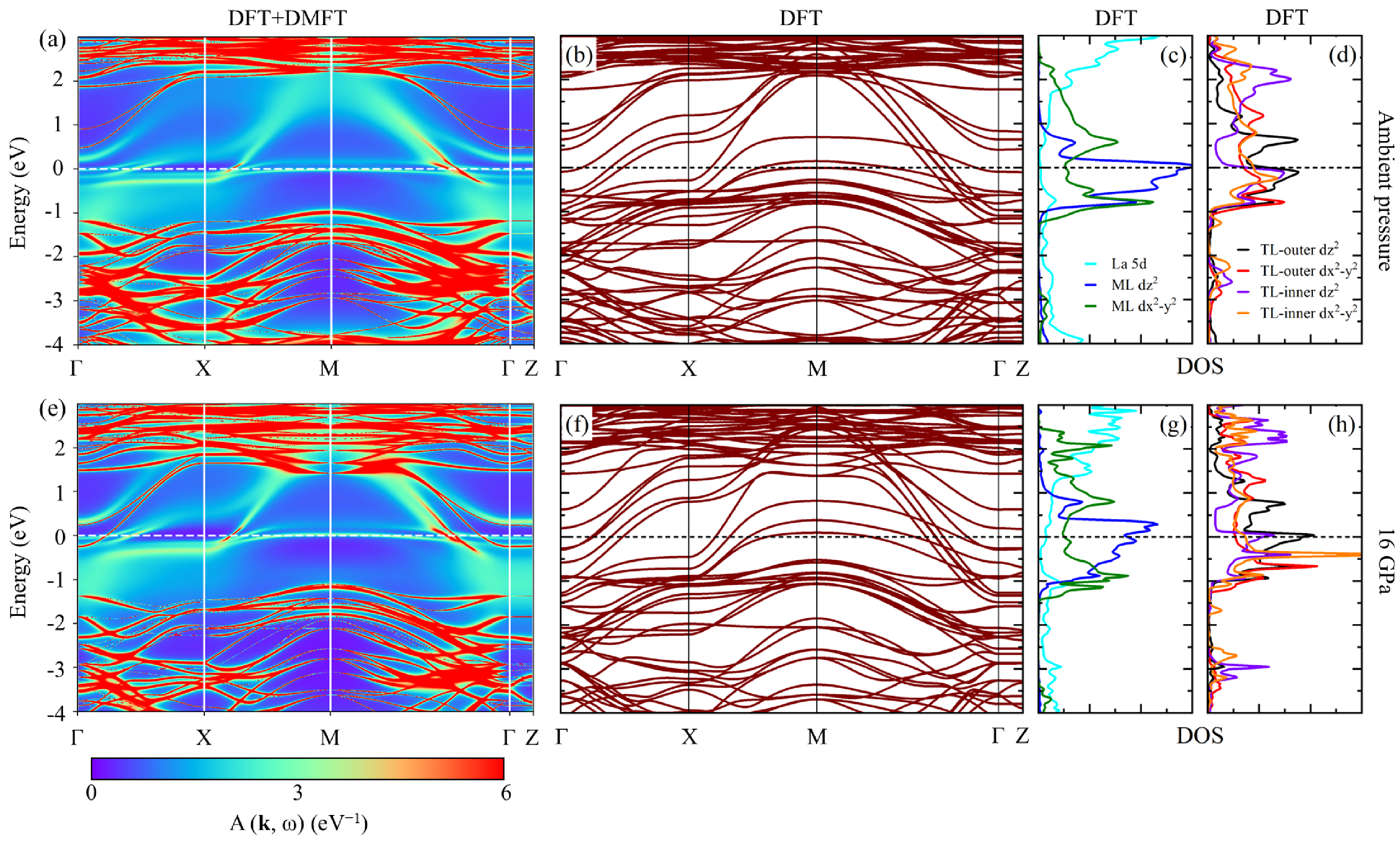}
\caption{Momentum-resolved spectral functions $A(\bf{k}, \omega)$ obtained by DFT+DMFT calculations at 290 K, DFT band structure, and orbital-projected density of states of 1313-La327 under ambient pressure (a-d) and 16 GPa (e-h), respectively. The Fermi level is set to zero.}
\label{fig:band}
\end{figure*}

However, a hidden phase of La$_3$Ni$_2$O$_7$ was experimentally observed recently~\cite{Chen-JACS2024}. Distinguished from the previously reported 2222-La327, this hidden phase exhibits a distinctive long-range structural order, stacking with monolayer and trilayer of NiO$_6$ blocks in a ``1313" sequence. And this structure has been subsequently reported by some other groups~\cite{puphal2023unconventional,Wang-InC,abadi2024electronic}. Among them, Puphal $\emph{et al.}$~\cite{puphal2023unconventional} claimed that they obtained single crystals of 1313-La327 as primary bulk phase and superconductivity was observed under high pressures. Other groups obtained samples including other competing phases such as 2222-La327, La4310, or La$_2$NiO$_4$~\cite{Wang-InC,abadi2024electronic}. These findings bring some controversy to the origin of superconductivity in La$_3$Ni$_2$O$_7$ compounds. Hence, it is urgent to study the correlated electronic structure in 1313-La327, which may provide clues about the possible superconductivity.

In this paper, we perform the density functional theory plus dynamical mean-field theory (DFT+DFMT) calculations to study the correlated electronic structure in 1313-La327. Our calculated Fermi surfaces are in good agreement with the angle-resolved photoemission spectroscopy (ARPES) results. We find that 1313-La327 is a multiorbital correlated metal and a Lifshitz transition is observed under 16 GPa. Furthermore, our results suggest that 1313-La327 is a unique platform, which may include Mott and Hund physics. Specifically, the calculated self-energy functions and orbital-resolved spectral functions show that the ML Ni-3$d_{z^2}$ orbitals exhibit a Mott behavior, while the ML Ni-3$d_{x^2-y^2}$ orbitals are metallic due to the self-doping between the ML and TL. An orbital-selective Mott phase (OSMP) is found in the ML. Besides, we find large weights of high-spin states in TL-outer Ni and ML Ni, which indicates Hundness. The calculated static local spin susceptibility suggests that there is a complicated magnetic property in 1313-La327.

\section{Results and analysis}

\subsection{Electronic structure}

We first focus on the momentum-resolved spectral functions $A(\bf{k}, \omega)$ of 1313-La327. As shown in Figs.~\ref{fig:band}(a) and (e), we find that the bands around the Fermi level are strongly renormalized. Especially for those bands near the Fermi level, the bandwidth is $\sim$ 0.5 eV. At the $M$ point, those bands around the Fermi level are almost invisible compared and the spectral weights reduce obviously because of electronic correlation. All these findings are similar to those in 2222-La327~\cite{Ouyang-prb2024} and La4310~\cite{Wang-prb2024}. These suggest that 1313-La327 is a correlated metal.

In order to make clear the characters of bands and the renormalization effect, we show the DFT calculated band structure and orbital-projected density of states (DOS) of 1313-La327 under ambient and 16 GPa pressures in Figs.~\ref{fig:band}(b-d) and (f-h). The DFT results suggest that 1313-La327 is a multiorbital metal, where the $e_g$ orbitals of all three kinds of  Ni atoms contribute the bands near the Fermi level. And the bandwidth of the Ni-$e_g$ orbitals is $\sim$ 4 eV, which is consistent with other previous theoretical reports~\cite{Chen-JACS2024,zhang2024electronicstructureselfdopingsuperconducting}. Furthermore, we find that at ambient pressure La-5$d$ orbitals contribute little to the DOS around the Fermi level [Fig.~\ref{fig:band}(c)]. However, under 16 GPa pressure the DOS of the La-5$d$ orbitals becomes finite at the Fermi level [Fig.~\ref{fig:band}(g)]. And a La-related band crosses the Fermi level around the $\Gamma$ point, which is mainly contributed by the surface La atoms of the ML and TL.

\begin{figure}[tbhp]
\centering
\includegraphics[width=8.6cm]{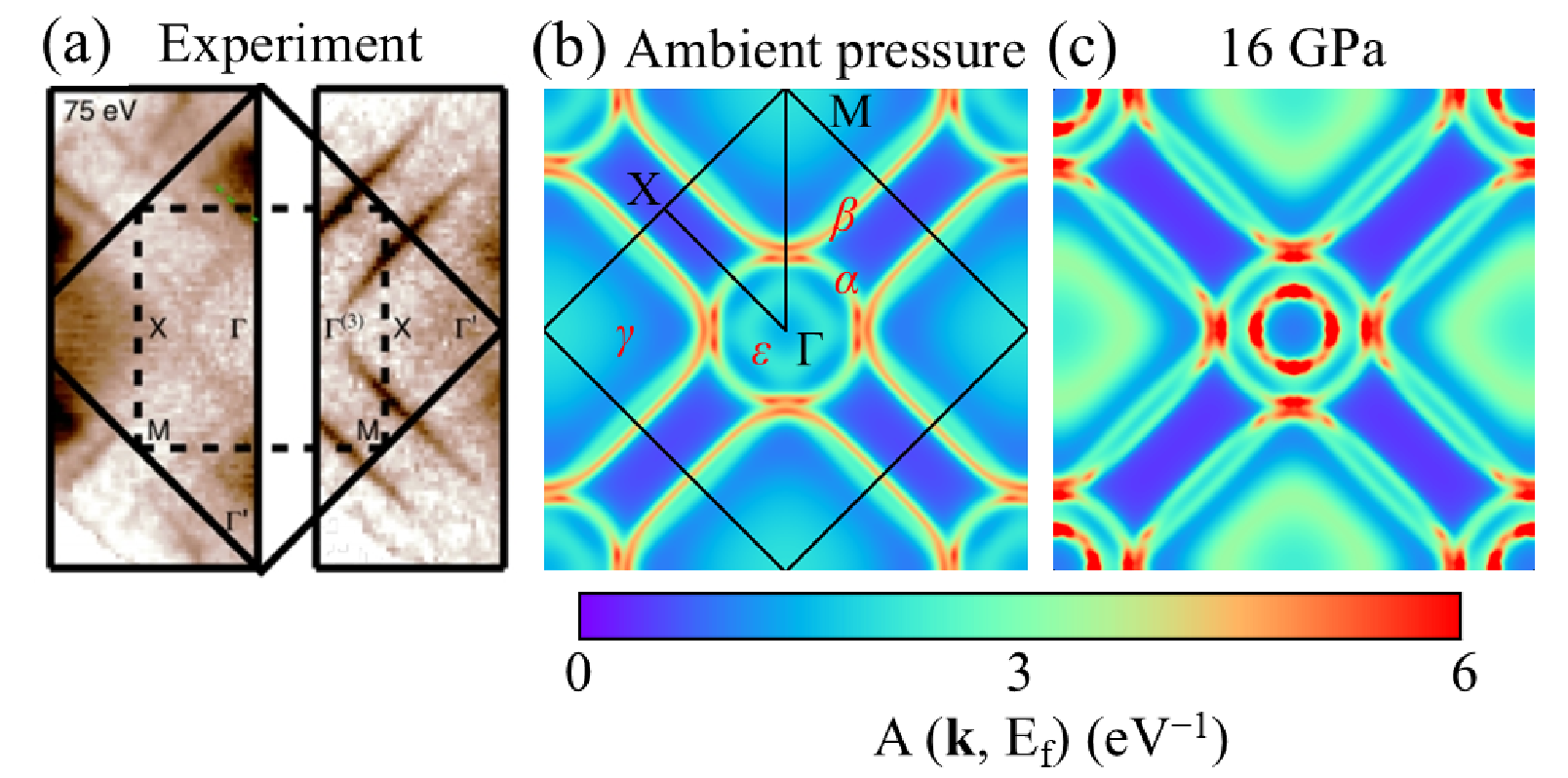}
\caption{(a) The Fermi surfaces at $k_z$ = 0 determined by the ARPES at ambient pressure from Ref.~\cite{abadi2024electronic}. The DFT+DMFT calculated Fermi surfaces at $k_z$ = 0 plane for ambient pressure (b) and 16 GPa (c) at 290 K, respectively. The first Brillouin zone and the high-symmetry paths are denoted by black solid lines in (b). }
\label{fig:FS}
\end{figure}

In Figs.~\ref{fig:FS}(b) and (c), we show the DFT+DMFT calculated Fermi surfaces at $k_z$ = 0 plane under ambient pressure and 16 GPa pressure, respectively. And the ARPES Fermi surfaces under ambient pressure~\cite{abadi2024electronic} are exhibited in Fig.~\ref{fig:FS}(a). Our calculated Fermi surfaces under ambient pressure are in good agreement with the ARPES results. Besides the $\alpha$, $\beta$, and $\gamma$ pockets which are similar to those in 2222-La327, a newly discovered $\epsilon$ pocket in 1313-La327 has also been well reproduced in our theoretical calculations. This $\epsilon$ pocket shows a diamond-like pattern around the $\Gamma$ point in Fig.~\ref{fig:FS}(b). However, in Fig.~\ref{fig:FS}(c), under 16 GPa pressure, it is clearly seen that there is a ring-like Fermi surface with large quasi-particle weight, which is quite close to the $\epsilon$ pocket. This phenomenon has not been found in our previous studies about 2222-La327~\cite{Ouyang-prb2024} or La4310~\cite{Wang-prb2024}. For example, the La-related bands in 2222-La327 have no contribution to the Fermi surfaces. But it is reminiscent of the self-doping in infinite-layer nickelates La$_{1-x}$Sr$_{x}$NiO$_2$, where La-related band crosses the Fermi level~\cite{Gu-Innovation3,Yang-FP9}.

\begin{figure*}[bthp]
\centering
\includegraphics[width=17.2cm]{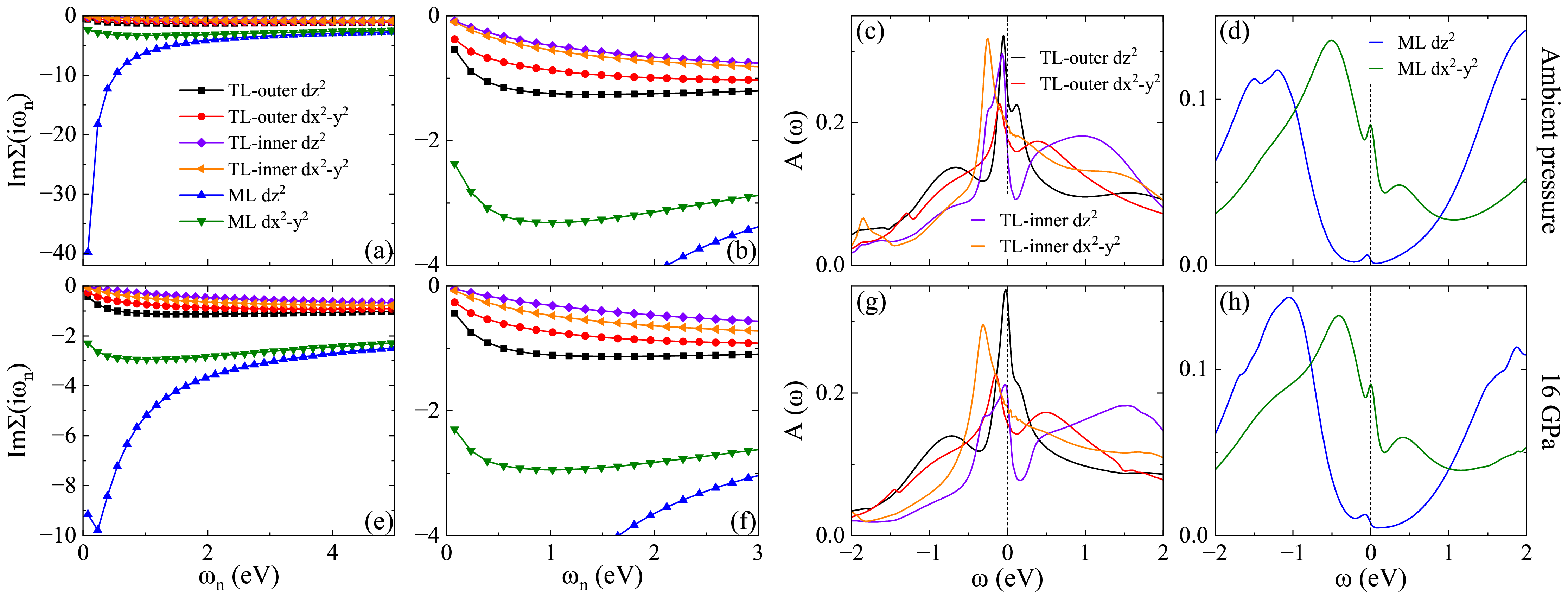}
\caption{The imaginary parts of Matsubara self-energy functions Im$\Sigma(i\omega_n)$ and orbital-resolved spectral functions $A(\omega)$ of 1313-La327 obtained by DFT+DMFT calculations at 290 K under ambient pressure (a-d) and 16 GPa (e-h). }
\label{fig:Fig3}
\end{figure*}

In Fig.~\ref{fig:Fig3}, we show the orbital-resolved Matsubara self-energy functions Im$\Sigma(i\omega_n)$ and local spectral functions $A(\omega)$ of 1313-La327 under ambient pressure and 16 GPa, respectively. Under ambient pressure, an obvious orbital difference is found in Fig.~\ref{fig:Fig3}(a). The self-energy function Im$\Sigma(i\omega_n)$ of the ML Ni-3$d_{z^2}$ orbital exhibits a divergent behavior in low-frequency region, which is totally different from the others. This is a signal of Mott physics. A Mott gap is observed in the local spectral function $A(\omega)$ of the ML Ni-3$d_{z^2}$ orbital in Fig.~\ref{fig:Fig3}(d). As for the ML Ni-3$d_{x^2-y^2}$ orbital, a large intercept of low-frequency Im$\Sigma(i\omega_n)$ in Fig.~\ref{fig:Fig3}(b) indicates its strong electronic correlation. It should be noted that both the ML Ni-3$d_{z^2}$ and 3$d_{x^2-y^2}$ orbitals show a small quasi-particle resonance peak at the Fermi level in our room-temperature calculations [Fig.~\ref{fig:Fig3}(d)]. Anyway, the characteristic of the energy gap is clear. Under 16 GPa, similar behaviors are observed. In Fig.~\ref{fig:Fig3}(e), a large and finite intercept of Im$\Sigma(i\omega_n)$ of the ML Ni-3$d_{z^2}$ orbital appears. But the local spectral function $A(\omega)$ of the ML Ni-3$d_{z^2}$ orbital still shows a Mott behavior [Fig.~\ref{fig:Fig3}(h)], which is similar to the ambient case. As for the TL Ni-$e_g$ orbitals, the Im$\Sigma(i\omega_n)$ of the TL-outer Ni-$e_g$ orbitals at low-frequency region show nonlinear behaviors. And finite intercepts of Im$\Sigma(i\omega_n)$ can be found around zero frequency. These suggest that the TL-outer Ni-$e_g$ orbitals show non-Fermi liquid behaviors. But the Im$\Sigma(i\omega_n)$ of the TL-inner Ni-$e_g$ orbitals show linear behaviors at low frequencies, which is consistent with a Fermi liquid behavior [Figs.~\ref{fig:Fig3}(b) and (f)]. Similar results have also been found in La4310~\cite{Wang-prb2024}.

 \subsection{Spin susceptibility}

 \begin{table}[bp]
\begin{center}
\renewcommand\arraystretch{1.5}
\caption{The DFT+DMFT calculated static local spin susceptibility $\chi_{\rm{spin}}$ of TL-outer Ni, TL-inner Ni, and ML Ni at 290 K under ambient pressure (AP) and 16 GPa, respectively. The $\chi_{\rm{spin}}$ of La4310 (44.3 GPa)~\cite{Wang-prb2024} is also exhibited.}
\label{tab1}
\begin{tabular*}{8.6cm}{@{\extracolsep{\fill}} cccc}
\hline\hline

    &TL-outer Ni     & TL-inner Ni   &ML Ni    \\

\hline

1313-La327 (AP)$\hspace{1.65em}$      &6.99    &1.36     & 27.72   \\

\hline

1313-La327 (16 GPa)      &5.34    &0.87     & 25.66   \\

\hline

$\hspace{3.35em}$ La4310 (44.3 GPa)      &1.92   &0.98     & $-$   \\

\hline\hline
\end{tabular*}
\end{center}
\end{table} 
 
According to the above results, a significant difference of physical properties between the TL and ML in 1313-La327 is observed. In Table~\ref{tab1}, we show the DFT+DMFT calculated static local spin susceptibility $\chi_{\rm{spin}}$ of the three kinds of Ni in 1313-La327, which further shows the difference between the TL and ML. The spin susceptibility of the ML Ni is much larger than that of the TL Ni, which suggests a local moment may exist on the ML Ni. Although there lack experimental reports about magnetic order in 1313-La327, this finding is very easy to associate with La$_2$NiO$_4$, which shares the same chemical formula with the ML in 1313-La327. It is known that the undoped La$_2$NiO$_4$ is an AFM Mott insulator. When applying hole doping, an insulator-metal transition occurs~\cite{Cava-prb1991,Chen-prl1993,Cheong-prb1994,Wu-prb2003}. In 1313-La327, the average chemical valence of Ni cation is Ni$^{2.5+}$. The ML and TL have chemical formulas of La$_2$NiO$_4$ and La$_4$Ni$_3$O$_{10}$, respectively, where the average chemical valence of Ni cations are Ni$^{2+}$ and Ni$^{2.67+}$, respectively. Hence, in order to make the chemical valence of the ML Ni and TL Ni cations get closer to Ni$^{2.5+}$, it is natural to speculate that there is a self-doping between the ML and TL, where electrons transfer from ML to TL and holes are doped in ML. And according to Refs.~\cite{Chen-JACS2024,puphal2023unconventional}, the in-plane Ni-O distance is shorter than the out-of plane Ni-O distance in 1313-La327, so the self-doping affects the high-lying Ni-3$d_{x^2-y^2}$ orbitals. This is consistent with our result in Figs.~\ref{fig:Fig3}(d) and (h), where the ML Ni-3$d_{x^2-y^2}$ orbital exhibits a metallic behavior. As for the TL Ni in 1313-La327, the $\chi_{\rm{spin}}$ of TL-outer Ni is larger than that of TL-inner Ni. All these findings suggest that there may be a more complicated magnetic property in 1313-La327, which is vital for superconducting pairing and needs more experimental and theoretical investigations.

 \begin{table*}[t]
\begin{center}
\renewcommand\arraystretch{1.5}
\caption{The DFT+DMFT calculated mass enhancement $m^*/m$ and the local occupation number $N_{e_g}$ of Ni-$e_g$ orbitals in 1313-La327 at 290 K under ambient pressure (AP) and 16 GPa and La4310 (44.3 GPa)~\cite{Wang-prb2024}, respectively.}
\label{tab2}
\begin{tabular*}{12cm}{@{\extracolsep{\fill}} rccccccc}
\hline\hline

\multicolumn{2}{c} {\multirow{2}*{{Orbital}}}                 & \multicolumn{2}{c}{TL-outer}     & \multicolumn{2}{c}{TL-inner}    &\multicolumn{2}{c}{ML}    \\
                                                                                         &     &3$d_{z^2}$    & 3$d_{x^2 - y^2}$    & 3$d_{z^2}$   & 3$d_{x^2 - y^2}$   &3$d_{z^2}$    &3$d_{x^2 - y^2}$    \\

\hline

\multirow{3}*{$m^*/m$}         &{1313-La327 (AP)$\hspace{1.5em}$}   & 5.374      & 4.098       & 2.255       & 2.264    & $-$   & $-$   \\
                                                   &{1313-La327 (16GPa)}   & 4.345      & 2.966       & 1.684       & 1.508    & $-$   & $-$   \\
                                                   &{$\hspace{3.8em}$La4310 (44.3 GPa)}  & 2.578      & 2.296       & 2.018       & 1.986    &$-$   &$-$   \\

\hline

\multirow{3}*{$N_{e_g}$}         &{1313-La327 (AP)$\hspace{1.5em}$}   & 1.115      & 1.049       & 1.021       & 1.043       &1.082   &1.083  \\
                                                   &{1313-La327 (16GPa)}   & 1.110      & 1.053       & 0.962       & 1.063       &1.104   &1.083  \\
                                                   &{$\hspace{3.8em}$La4310 (44.3 GPa)}  & 1.088      & 1.060       & 1.060       & 1.017       &$-$   &$-$  \\







\hline\hline
\end{tabular*}
\end{center}
\end{table*}

 \begin{table*}[thbp]
\begin{center}
\small
\renewcommand\arraystretch{1.5}
\caption{The DFT+DMFT calculated weights (\%) of the Ni-$e_g$ orbital local multiplets in 1313-La327 at 290 K under ambient pressure (AP) and 16 GPa, La4310 (44.3 GPa)~\cite{Wang-prb2024}, and 2222-La327 (29.5 GPa)~\cite{Ouyang-prb2024}, respectively. The good quantum numbers $N_{\Gamma}$ and $S_z$ denote the total occupancy and total spin of the Ni-$e_g$ orbitals, which are used to label different local spin states.}
\label{tab3}
\begin{tabular*}{12cm}{@{\extracolsep{\fill}} rccccccc}
\hline\hline

\multicolumn{2}{c} {$N_{\Gamma}$}   & 0   & 1   & 2   & 2   & 3   & 4   \\

\multicolumn{2}{c} {$S_z$}  & 0    & 1/2   & 0   & 1   & 1/2   & 0  \\

\hline

\multicolumn{2}{c} {$\hspace{7em}$2222-La327 (29.5 GPa)}    & 0.4   & 12.4   & 23.6   & 33.3   & 28.1  & 2.3  \\

\hline

\multirow{3}*{TL-outer Ni} &1313-La327 (AP)$\hspace{1.7em}$   & 0.3   & 11.4   & 19.9  & 41.7   & 25.1  & 1.6  \\
                                                &1313-La327 (16 GPa)  &0.4   & 12.5   & 21.7   & 37.7   & 25.9  & 1.9  \\
                                                &$\hspace{3.55em}$La4310 (44.3 GPa)   & 0.6   & 14.7  & 24.8   & 31.5   & 26.1  & 2.1  \\
                                                       
\hline

\multirow{3}*{TL-inner Ni} &1313-La327 (AP)$\hspace{1.7em}$  &0.9   & 18.1   & 26.9   & 29.3   & 23.1  & 1.6  \\
                                                              &1313-La327 (16 GPa)   & 1.3   & 20.9   & 28.1   & 25.5   & 22.5  & 1.8  \\ 
                                                              &$\hspace{3.55em}$La4310 (44.3 GPa)   & 1.0   & 18.3   & 27.1   & 27.6   & 24.1  & 1.9  \\

\hline

\multirow{2}*{ML Ni} &1313-La327 (AP)$\hspace{1.7em}$   & 0.0   &$\hspace{0.5em}$3.3  & $\hspace{0.5em}$4.7   & 72.8   & 18.7  & 0.5  \\
                                                              &1313-La327 (16 GPa)   & 0.0   & $\hspace{0.5em}$3.8   & $\hspace{0.5em}$5.9   & 68.5   & 21.1  & 0.8  \\

\hline\hline
\end{tabular*}
\end{center}
\end{table*}

\subsection{Electronic correlation}

 Here, we investigate the electronic correlation in 1313-La327 and make systematic comparisons with those previous nickelate superconductors, for example 2222-La327~\cite{Ouyang-prb2024} and La4310~\cite{Wang-prb2024}. In Table~\ref{tab2}, we show the mass enhancement $m^*/m$ and the local occupation number $N_{e_g}$ of 1313-La327 and La4310. The mass enhancement is defined as
\begin{equation}
{m^*}/m = {Z^{-1}} = 1 - \frac{\partial \mbox{Re} \Sigma (\omega)}{\partial \omega} \bigg|_{\omega = 0}.
\label{eq:massenhancement}
\end{equation}     
$Z$ is the renormalization factor, which is a key factor to characterize band renormalization and electronic correlation. As for the TL in 1313-La327 under ambient pressure, the mass enhancement of TL-outer Ni-$e_g$ orbitals is larger than that of the inner Ni-$e_g$ orbitals, which suggests that the TL-outer Ni-$e_g$ orbitals are more correlated than the TL-inner Ni-$e_g$ orbitals. Furthermore, we find that the occupation of the TL-inner Ni-$e_g$ orbitals is less than that of the TL-outer Ni-$e_g$ orbitals (Table~\ref{tab2}), which suggests that the TL-inner layer has more hole doping and hence possesses weaker electronic correlation. Similar findings have been discussed in La4310~\cite{Wang-prb2024}. In addition, we also find that pressure weakens the electronic correlation in 1313-La327, and the $e_g$ orbitals of TL-outer Ni possess stronger renormalization than that of TL-inner Ni. 

As for the local occupation number $N_{e_g}$ of the Ni-$e_g$ orbitals in 1313-La327, due to orbital hybridization, our calculated $N_{e_g}$ is larger than the nominal value estimated from chemical valence. It should be noted that a strong $d$-$p$ hybridization between the Ni-3$d_{z^2}$ and inner apical O-2$p_z$ orbitals, as well as between the Ni-3$d_{x^2-y^2}$ and in-plane O-2$p_{x}/p_{y}$ orbitals in 2222-La327 have been experimentally confirmed recently~\cite{Dong-nature}. The holes transfer from Ni cations to O anions, hence the exact local occupation number of the Ni-3$d$ orbitals is larger than the nominal value of 7.5. This result shows good consistency with our previous theoretical calculations about 2222-La327~\cite{Ouyang-prb2024} and La4310~\cite{Wang-prb2024}. And a previous theoretical report also predicted an approximate Ni-3$d^8$ occupation in 1313-La327~\cite{lechermann2024electronic}. Our finding about the local occupation number of the Ni-$e_g$ orbitals in 1313-La327 once again suggests that strong $d$-$p$ hybridization may be a common feature in these RP series of nickelates.

 \subsection{Hundness from high-spin state}
 
In Table~\ref{tab3}, we show the weights of the Ni-$e_g$ orbitals local multiplets in 1313-La327 and make comparisons with those in 2222-La327 and La4310. According to our previous study, 2222-La327 is a Hund metal and the high-spin $S_z$ = 1 state possesses a larger weight~\cite{Ouyang-prb2024}. And Hund correlation plays an important role in superconducting pairing in 2222-La327. Specifcally, the AFM interactions between the interlayer Ni-3$d_{z^2}$ orbitals are shared to the Ni-3$d_{x^2-y^2}$ orbitals by Hund's correlation. Then these AFM interactions provide glue for the superconducting pairing~\cite{Chen-prl2024,Tian-prb2024,Oh-prb2023}. Besides, Hund's correlation has also been found in infinite-layer LaNiO$_2$~\cite{Wang-PRB102}. Here, we find large weights of high-spin states of the TL-outer Ni and ML Ni in 1313-La327. These findings suggest that Hund's correlation may be a common property in nickelates. In addition, it should be noticed that the high-spin $S_z$ = 1 state of ML Ni has a very large weight of $\sim$ 70\%. This finding supports the previous analysis of the results about $\chi_{\rm{spin}}$ in Table~\ref{tab1} and once again emphasizes that there may be a complex magnetic property in 1313-La327.

\subsection{Results for low and zero temperatures}
 
 \begin{figure}[b]
\centering
\includegraphics[width=8.6cm]{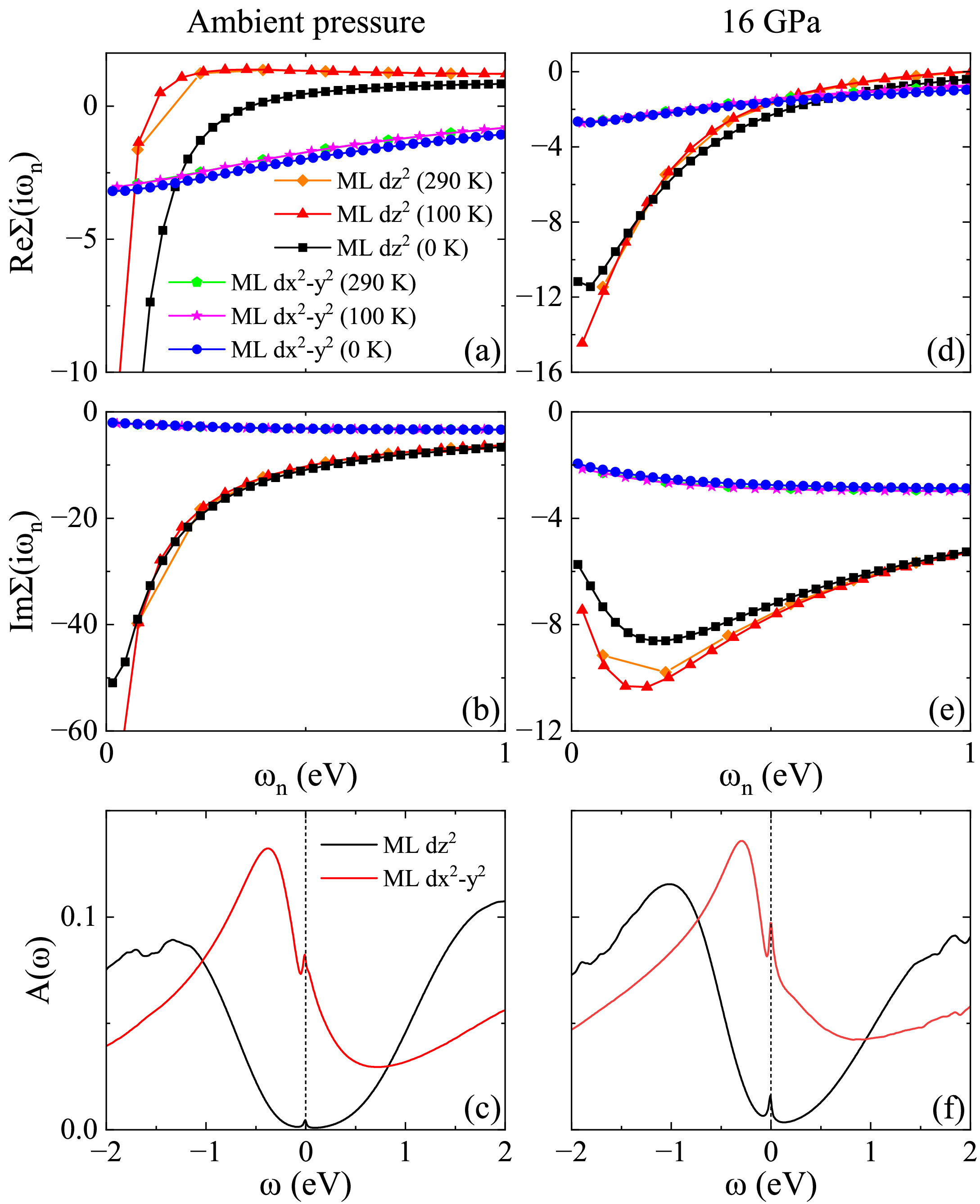}
\caption{Matsubara self-energy functions of 1313-La327 at 290, 100, and 0 K and spectral functions $A(\omega)$ at 0 K under ambient pressure (a-c) and 16 GPa (d-f), respectively. The zero-temperature results are obtained by the NORG impurity solver~\cite{PhysRevB.89.085108}.}
\label{fig:Fig4}
\end{figure}
 
In order to further discuss the orbital-selective Mott phase in the ML Ni-$e_g$ orbitals and describe the low-temperature physics, we perform zero-temperature DFT+DMFT calculations with natural orbitals renormalization group (NORG) impurity solver and make comparisons with results of finite temperatures.

As shown in Figs.~\ref{fig:Fig4}(a-b) and (d-e), the zero-temperature Matsubara self-energy functions exhibit similar behaviors to those of finite temperatures. The Im$\Sigma(i\omega_n)$ of the ML Ni-3$d_{{x^2}-{y^2}}$ orbital have a finite intercept under both ambient and 16 GPa pressures, showing a non-Fermi liquid behavior. But the Im$\Sigma(i\omega_n)$ of the ML Ni-3$d_{z^2}$ orbital at ambient pressure show a divergent behavior all along at low-frequency region as temperature decreases, which indicates Mott physics. Under 16 GPa pressure, the finite intercepts of the Im$\Sigma(i\omega_n)$ of the ML Ni-3$d_{z^2}$ orbital indicate a non-Fermi liquid behavior.

In Figs.~\ref{fig:Fig4}(c) and (f), we show the spectral functions $A(\omega)$ of the ML Ni-$e_g$ orbitals at zero temperature, where the quasi-particle resonance peaks at the Fermi level remain. And this situation is well consistent to an OSMP scenario. Specifically, the ML Ni-3$d_{z^2}$ and Ni-3$d_{x^2-y^2}$ orbitals would be insulating and metallic, respectively, if there was no interorbital hopping. The appearance of quasi-particle peaks at the Fermi level is attributed to the finite interorbital hopping between the ML Ni-3$d_{z^2}$ and Ni-3$d_{x^2-y^2}$ orbitals. And a previous theoretical report regarding 2222-La327 suggests that there is a finite effective interorbital hopping between the Ni-3$d_{z^2}$ and Ni-3$d_{x^2-y^2}$ orbitals~\cite{PhysRevLett.131.126001}. The OSMP in 1313-La327 is a realistic example of theoretical report~\cite{PhysRevLett.129.096403}. The $A(\omega)$ of the heavy orbital develops a thin quasi-particle peak at the Fermi level. This OSMP scenario is reminiscent of Kondo physics where hybridization happens between local moments and conducting electrons. But here the hybridization is between local moments (here ML Ni-3$d_{z^2}$ orbitals)  and a non-Fermi liquid (here ML Ni-3$d_{x^2-y^2}$ orbitals).

\section{Discussion and Conclusion}
Although the superconducting transition in 1313-La327 still needs to be further experimentally confirmed, our theoretical calculations indeed reveal some unique features. 

(i) We find a self-doping in 1313-La327 between ML and TL. Note that this should not be confused with the self-doping revealed in 2222-La327 from a recent experiment~\cite{Dong-nature}, where the self-doped ligand holes mainly resides on the inner apical O and in-plane O atoms. 

(ii) The interorbital hopping between the localized ML Ni-3$d_{z^2}$ and delocalized Ni-3$d_{x^2-y^2}$ orbitals breaks the OSMP of the ML, giving rise to quasi-particle peaks in the spectra of the ML Ni-$e_g$ orbitals at the Fermi level. This suggests that 1313-La327 is a  realistic example of the OSMP scenario~\cite{PhysRevLett.129.096403}. 

(iii) Some previous works claimed that doping electrons is beneficial for achieving a higher superconducting $T_c$ in La4310~\cite{zhang2-arXiv,lu2024superconductivity}. From this aspect, the self-doping provides a potential support for the possible realization of a higher superconducting $T_c$ in the TL of 1313-La327. This may explain the high-$T_c$ superconducting transition under pressures, which was reported by Puphal $\emph{et al.}$~\cite{puphal2023unconventional}. 

(iv) A Lifshitz transition is found in 1313-La327 under 16 GPa and a La-related band crosses the Fermi level. This is similar to the self-doped infinite-layer La$_{1-x}$Sr$_x$NiO$_2$~\cite{Gu-Innovation3,Yang-FP9}.  

(v) A very large Im$\Sigma(i\omega_n)$ of the ML Ni-3$d_{x^2-y^2}$ orbitals and nonlinear Im$\Sigma(i\omega_n)$ of the TL-outer Ni-$e_g$ orbitals indicate their strong correlation [Figs.~\ref{fig:Fig3}(a-b) and (e-f)]. Refs.~\cite{Wang-prb2024} and~\cite{Tian-prb2024} show that Hund's correlation can significantly enhance electronic correlation and give rise to a non-Fermi liquid behavior. Here for 1313-La327, the Hund's correlation is manifest through Ni local high-spin state and should account for the strong correlation as well as the non-Fermi liquid behavior. And we suggest that Hund's correlation may be a common property of nickelates.

All these should be further both experimentally and theoretically investigated.

In Summary, we have systematically studied the correlated electronic structure in the newly discovered 1313-La327. Our calculated Fermi surfaces are in good agreement with the ARPES results. We find that 1313-La327 is a multiorbital correlated metal. Due to the self-doping from the ML to TL, the ML Ni-3$d_{x^2-y^2}$ orbitals are metallic and show a non-Fermi liquid behavior. But the ML Ni-3$d_{z^2}$ orbitals show a Mott behavior. And the interorbital hopping between the localized ML Ni-3$d_{z^2}$ and delocalized Ni-3$d_{x^2-y^2}$ orbitals causes an OSMP. Our calculations about the static local spin susceptibility suggest that there may be large spin fluctuation in ML Ni and 1313-La327 may have a more complicated magnetic property. As for the TL in 1313-La327, the TL-outer Ni-$e_g$ orbitals show a non-Fermi liquid behavior, which is similar to the case of La4310. The large weights of high-spin states of ML Ni and TL-outer Ni indicate Hundness in 1313-La327. Under 16 GPa pressure, a Lifshitz transition is observed and a La-related band crosses the Fermi level. Our studies give a new perspective to understand this ML-TL stacking structure and provide a theoretical reference for studying other potential mixed-stacked nickelate superconductors.

\begin{acknowledgments}
This work was supported by the National Natural Science Foundation of China (Grant No. 11934020).
Z.Y.L. was also supported by Innovation Program for Quantum Science and Technology (Grant No. 2021ZD0302402).
Computational resources were provided by the Physical Laboratory of High Performance Computing at Renmin University of China.
\end{acknowledgments}

\appendix

\section{DETAILS OF THE METHOD}
The DFT parts of our DFT+DMFT calculations were performed by WIEN2K code with the full-potential linearized augmented plane-wave method~\cite{Blaha-JCP152}. The experimental crystal structures under ambient pressure and high pressure were chosen in our calculations according to Refs.~\cite{Chen-JACS2024} and~\cite{puphal2023unconventional}, respectively. The cutoff parameter was $R_{\rm{MT}}K_{\rm{max}}$ = 7.0. The generalized gradient approximation with Perdew-Burke-Ernzerhof functional was chosen as the exchange and correlation potential~\cite{Perdew-PRL77}. The $k$-point mesh for the Brillouin zone integration was 13 $\times$ 13 $\times$ 2. The EDMFTF software package was used to perform the charge fully self-consistent DFT+DMFT calculations~\cite{Haule-PRB81}. Each DFT+DMFT cycle contained one-shot DMFT calculation and maximum 100 DFT iterations. Within about 40 DFT+DMFT cycles, we obtained a good convergence. The convergence criteria for charge and total energy were $10^{-7}$ eV and $10^{-7}$ Ry, respectively. The systems were enforced to be paramagnetic. Only Ni-$e_g$ orbitals were treated as correlated. A three-impurity problem was considered in our calculations because there are three different kinds of Ni atoms in 1313-La327. They are TL-outer Ni, TL-inner Ni, and ML Ni, respectively. The Hund's exchange parameter $J_H$ was set to be 1.0 eV. We chose the Coulomb interaction parameter $U$ = 5.0 (8.0) eV for the Ni atoms in TL (ML) structure. And the experimentally determined Fermi surfaces were well reproduced in our calculations. The density-density form of the Coulomb repulsion was used. We used the projectors with an energy window from $-$10 to 10 eV with respect to the Fermi level to construct the correlated orbitals. The finite-temperature and zero-temperature quantum impurity problems for the DMFT were respectively solved by the hybridization expansion continuous-time quantum Monte Carlo impurity solver~\cite{Haule-PRB75} and NORG impurity solver~\cite{PhysRevB.89.085108} with exact double-counting scheme~\cite{Haule-PRL115} for the self-energy function. Utilizing the maximum entropy method analytical continuation~\cite{Jarrell-PR269}, we obtained the real-frequency self-energy function, which was used to calculate the momentum-resolved spectral function and the other related physical quantities.

\bibliography {lno1313}

\end{document}